# SECURITY FRAMEWORK FOR IOT DEVICES AGAINST CYBER-ATTACKS

Aliya Tabassum[1] and Wadha Lebda[2]

Department of Computer Science and Engineering, Qatar University, Doha, Qatar
[1]`atabassum@qu.edu.qa` and [2]`wadha.lebda@qu.edu.qa`

## ABSTRACT

*Internet of Things (IoT) is the interconnection of heterogeneous smart devices through the Internet with diverse application areas. The huge number of smart devices and the complexity of networks has made it impossible to secure the data and communication between devices. Various conventional security controls are insufficient to prevent numerous attacks against these information-rich devices. Along with enhancing existing approaches, a peripheral defence, Intrusion Detection System (IDS), proved efficient in most scenarios. However, conventional IDS approaches are unsuitable to mitigate continuously emerging zero-day attacks. Intelligent mechanisms that can detect unfamiliar intrusions seems a prospective solution. This article explores popular attacks against IoT architecture and its relevant defence mechanisms to identify an appropriate protective measure for different networking practices and attack categories. Besides, a security framework for IoT architecture is provided with a list of security enhancement techniques.*

## 1. INTRODUCTION

In recent years, the number of smart IoT devices has increased dramatically. Due to the cheaper costs of hardware and open-source software, various companies are manufacturing IoT devices. A report published by HP, as a part of the Open Web Application Security Project (OWASP), proves that manufacturers ignore security aspects while developing these devices [1]. Hence, IoT devices have become potentially vulnerable targets for cybercriminals. In addition, it has become difficult for security specialists to secure the huge amount of data residing on the devices and the data in transmission in IoT networks. The complexity due to the number of IoT devices and networks provide opportunities to hackers to turn simple devices like TVs, cameras, DVDs and hubs into harmful botnets to launch jeopardizing cyberattacks [2]. To incorporate major security solutions such as cryptography in IoT devices there are two major challenges: (1) disestablished architecture, infrastructure and standards (2) unsupportive and insufficient resources. Applying appropriate defence mechanism (mitigation) is necessary to block the adversaries to reduce impact on the devices and/or end-users. Although the ever-increasing attacks are difficult to be mitigated fully, real-time network monitoring using an Intrusion Detection and/or Prevention system and adoption of strong access control & authentication mechanism can prevent attacks. The goal of our research is to provide detailed analysis of types of existing defence mechanisms for various attacks detection. So, that the most appropriate approach suitable to the current IoT networking is identified. In this paper, we explore persistent attacks against IoT devices and networks. After which, we provide details on current trends of security mechanisms that are being adopted to secure IoTs against such attacks. Further, we deduce the future deterministic metrics of IDS after a precise study of various IDS developments in literature. Lastly, from the analysis and review, we suggest a robust framework for securing IoT devices. The structure of the paper is as follows: section 2 provides background and overview on IoT devices, followed by the recurrent attacks against IoT architecture and various security mechanisms developed by security experts, in section 3. Section 4 elaborates on the types of significant security mechanisms that are potential in securing heterogeneous IoT networks. Later, section 5 recollects the crucial security mechanisms and a security framework. Finally, section 6 concludes the work.

## 2. BACKGROUND

IoT is an interconnection of billions of heterogeneous objects through the Internet. The number of connected smart IoT devices have surpassed the human population and in 2018, the number reached 7 billion. Moreover, researchers predict that in 2025 this number may peak to 22 billion with expected economy generated by various application domains is 4 to 11 Trillion Dollars [3] [4]. Figure 1 shows various application areas of IoT devices, which includes Smart Grid, Smart Retail, Smart Supply Chain, Smart Agriculture, Smart Industry, Smart Transportation, Smart Health, Smart Wearables, Smart Housing & Buildings and Smart City. From the mentioned statistics and areas of application, it is clear that IoTs are present in almost every sector and so, it has become essential to know how an IoT device works.

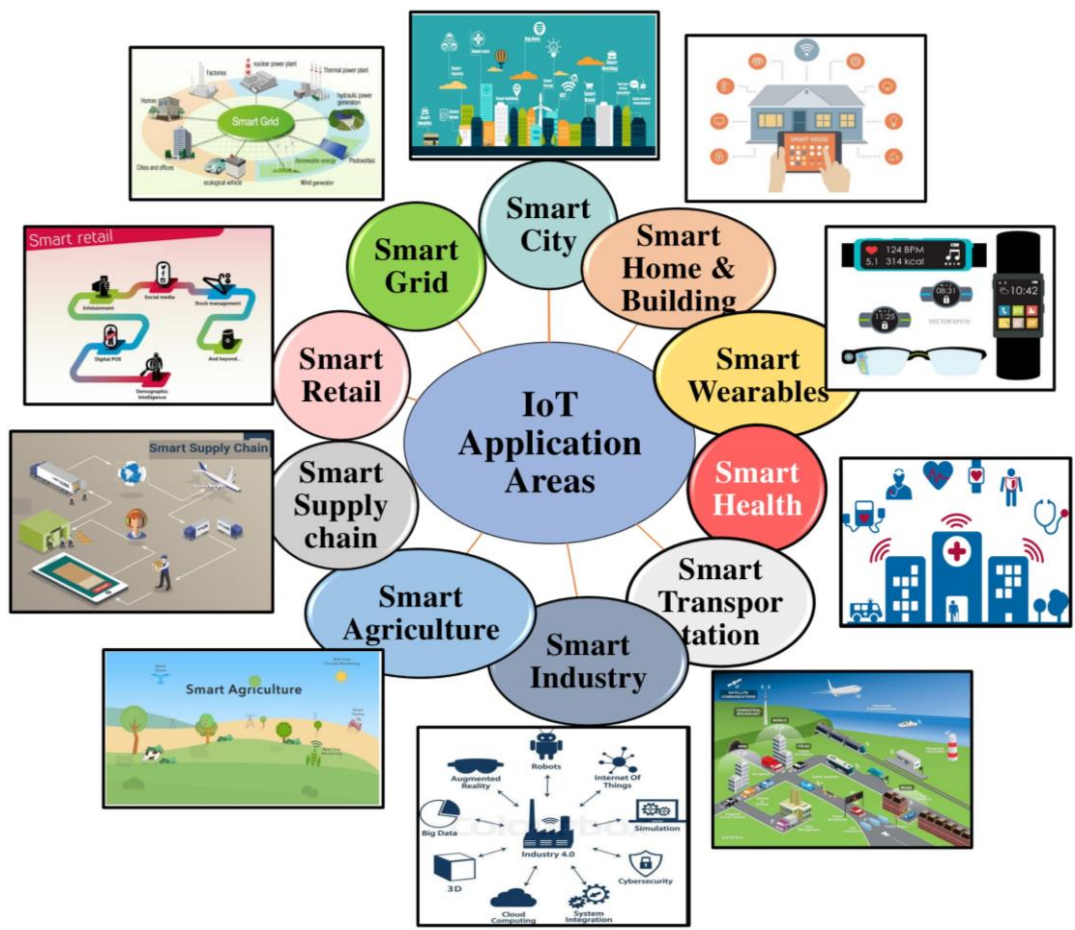

Figure 1. IoT application areas

Any IoT device operates in 3 phases: Collection Phase, Transmission phase and Processing, Management, Utilization phase.

- Collection Phase: It is the initial step to collect data from the physical environment using short-range communication sensing devices and technologies [5]. The devices for this phase have less battery power, limited memory and processing power. The design of the communication protocols is in such a way that it consumes less energy, operates on limited data rate, small memory and processing power for short distances. Because of the above reasons, these networks are referred to as Low-power and Lossy networks (LLNs). Consequently, the security mechanisms must be adaptable to the resource constraints of these devices.

- Transmission phase: This phase transmits the data collected from the Collection phase to the users and applications using transmission technologies such as Ethernet, Wi-Fi, Bluetooth, Bluetooth Low Energy (BLE), Hybrid Fiber Coaxial (HFC) and Digital Subscriber Line (DSL) [6]. Most of these technologies are vulnerable to attacks. Gateways integrate LLN protocols employed in the collection phase with the Internet protocols of transmission phase.

- Processing, Management, Utilization phase: The applications of this phase processes the collected data to get information about the environment. Sometimes, the applications have to make decisions based on the collected information [7]. It also has a middle-ware to integrate the communication with physical objects and multi-operation applications.

The above phases of operation need protection to ensure appropriate delivery of services. In the next section, we explore recurrent security attacks against IoT architecture.

## 3. ATTACKS AGAINST IoT LAYERS

Although there is no standardized model of IoT architecture, the basic types of architectures that are popularly used are 3 layers, 4 layer and 5 layer architectures and the recent advancements have more abstract layers added to these [8]. In our article, we explore the attacks in three layers, Perception, Transport and Network, shown in Figure 2 as these layers are highly targeted by security attacks.

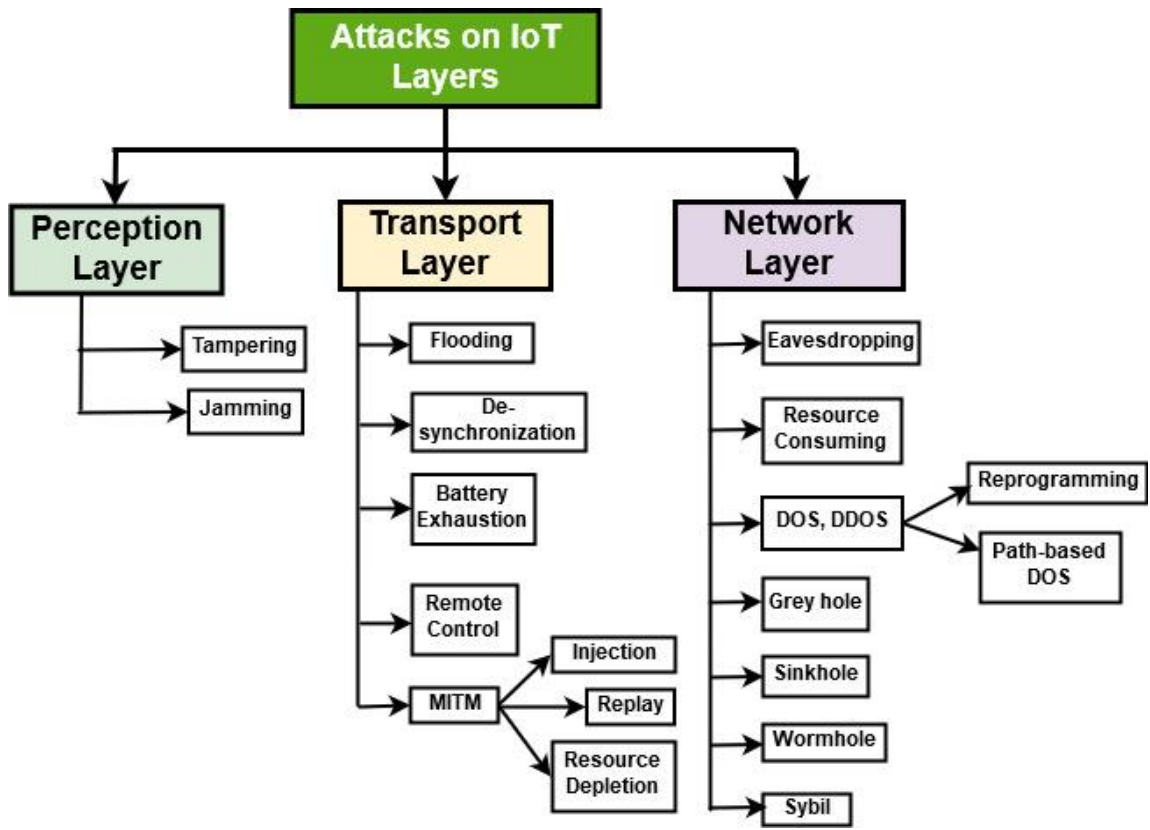

Figure 2. Attacks against IoT layers.

The types of attacks mentioned in Figure 2 are the ones, which are discussed by authors; there are other attacks, which have not been taken into consideration due non-popularity of the attacks or due to out of the scope of security mechanisms that are needed for such attacks.

## 3.1. Perception Layer

This is the first layer, which consists of the physical sensors and actuators of the IoT devices to sense the environment and collect information. The widespread attack at this layer is jamming and tampering. In a jamming attack, an adversary disrupts the operation of the network by squeezing/jamming the communication using high radio frequency signals [9]. Sometimes, an adversary can attack any sensor node to block the complete network resulting in a Denial of Service (DoS) or Distributed Denial of Service (DDoS) attack. Nowadays, cybercriminals use intelligent techniques to launch jamming attacks to evade various defensive measures like IDS/ Intrusion Prevention System (IPS). In defence to such attacks, a monitoring system is proposed by Liu et al. to distinguish interference and a real transmission where the energy consumed is verified each time to make sure it is not an attack [10]. This feature and energy monitoring system can identify channel interference efficiently but fall short for other attacks. Another model proposed using Monitor-Analyze-Plan-Execute (MAPE), which analysed signal strength but had similar drawback as of the previous one [11]. Multi-Agent Reinforcement Learning (MARL) algorithm is incorporated using Q-Learning to deal with jamming attacks and it gave 73% of performance. Likewise, an advanced Deep Learning (DL) framework is developed by researchers [12] to launch and mitigate jamming attacks. In this work, the Jammer senses the spectrum and if its classifier predicts any transmission to be successful, then the jammer blocks the transmission. Whereas, the defender system misleads the jammer decisions by propagating error signals. However, the success ratio of this model was very less & the maximum success ratio it gave was 69%. In both of these models, improved performance is indeed required to deal with real-time jamming attacks. From the discussion, we deduce that intelligent monitoring and learning models have the potential for detection of jamming attacks. Table 1 summarizes the discussed methods for a quick recap of the discussion.

Table 1. Perception layer attacks.

| Attack | Technique & Implications | Defence Mechanism |
| --- | --- | --- |
| WIRELESS JAMMING ATTACK | Jamming communication using high radio frequency signals [9]. | Distributing the usage across the spectrum and continuous monitoring of cognitive spectrum [11]. |
| | | Accessing received signal strength by using MAPE architecture [12]. |
| | Random and sensing-based jamming attacks using Deep Learning [12]. | Deep Learning framework to divert and corrupt the jammer decisions [12]. |
| | Complete jam of Wi-Fi signals and degradation of network performance | Reinforcement Learning for mitigating jamming based on Q-learning algorithm [13]. |

## 4.2. Transport Layer

This layer controls end-to-end links; and it mainly faces two types of attacks, flooding and the de-synchronization attacks. In flooding attack (TCP Synchronization / TCP-SYN), the memory resources of the devices are drained by propagating a control signal repetitively. Whereas, in the de-synchronized attack, the attacker interrupts a fully established communication link between two genuine end nodes by re-synchronization (infinite cycle) of their transmission. It disrupts the communication and exhaust resources of the network. Such type of attacks leads to altering and draining out the network performance. A mitigation system based on rate-limiting model in Contiki Operating System (OS) proves efficient to identify UDP Flood attacks [14] but fails to work well in TCP. Early detection modules of Flooding attacks are developed by using Software Defined Networking (SDN) but the model lacked practical testing in real-time scenarios [15] [16]. Table 2 elucidates other cyber-attacks against this layer along with the security measures.

1) Battery Exhaustion Attack: It occurs due to more consumption of power while processing the tasks such as transmitting, maintaining and receiving data. An attacker injects malicious processing codes to elongate the task, sometimes making the device ineffective. This attack is most popular in mobile devices. An IDS is proposed by Nash et.al [17] to overcome this attack. The system monitors the battery level of the device and it estimates the power requirements for each task. When the power consumed is greater than the threshold estimated, it triggers an alert terminates the task to avoid exhaustion of the battery. However, IDS designed on one / two features is not able to unmask other attacks and requires customization as per the attack.

2) Remote Control Attack: In this attack, the attacker tries to intercept communication between two parties by using botnets or Man-in-the-Middle (MITM) attacks to gain full control of the device [18]. In some cases, the attacker may launch a DoS attack to disrupt resources or the whole device. Such kind of attack may cause devastating implications in wearable sensors or Medical IoT devices. Researchers have developed approaches to protect against Remote control attack by incorporating Transport Layer Security (TLS) and Datagram Transport Layer Security (DTLS) for Constrained Application Protocol (CoAP) based LLNs [19]. CoAP is the widely used protocol in LLNs. Nevertheless, nowadays, many other protocols emerged for IoT networking such as MQ Telemetry Transport (MQTT), which consumes lesser energy of the devices. Hence, an approach has to be able to adapt to different protocols. Two other approaches shown in Table 2 are authentication and access control based, which are effective for Control Systems, Smart Grid, Home Automation and centralized control systems. However, is not effective for decentralized systems.

3) Man in the Middle (MITM) Attack: Weak security measures has given a stringent way for attackers to hold and vanish the resources of sensor devices. The unencrypted communication path is prone to attacks. An attacker can manipulate or delete information, violating the integrity, which may lead to various attacks: DoS, Eavesdropping, unauthorized access for tampering the data, injecting false information (authenticity) and Replay, Resource depletion and Injection attack [20]. In Eavesdropping, an adversary listens to the communication between the devices to know the capability and settings of the device to launch an attack. In a MITM attack, an attacker taps between two communicating devices by establishing a communication link and assuring them as authorized one by sending information to both and disconnecting their original communication link. It allows the intruder to acquire the user's data in an unethical way. The effective solutions for such attacks involve authentication and IDS system using Machine Learning, which give acceptable accuracy to defend against those attacks [21] [22].

Table 2. Transport layer attacks.

| Attack | Technique & Implications | Defence Mechanism |
|---|---|---|
| Flooding Attack, ICMP/ TCP/ UDP/ HTTP/ DNS | A repetitively propagating control signal drains memory and battery [9].<br><br>May lead to DDoS attack or jamming attack. | Rate-limiting mechanism in Contiki OS [14]. |
| | | SDN based IDS for monitoring activity [16]. |
| | | Dynamic Anomaly Detection module by learning attack behaviour [15]. |
| Battery Exhaustion Attack | Malicious codes to elongate the tasks & consumes more power, sometimes makes the device ineffective [17]. | IDS monitors the power consumed for tasks. If greater than the threshold estimated, an alert is triggered [17]. |
| Remote Control Attack | Intercepts communication using botnets or MITM attacks [18] to gain full control of the device and to disrupt resources of the whole device.<br><br>Devastating in smart home, ICS, smart grid, power and energy management systems. | TLS & DTLS security model for CoAP based LLNs [19]. |
| | | Identity monitoring system for ICS using cryptography, image processing, authentication and authorization [23]. |
| | | Multi-path onion IoT gateways, hidden IoT nodes using Tor services making them accessible to only authorized users [24]. |
| Man In The Middle (MITM) Attack | Attacker taps to manipulate or delete information. It can lead to DoS, replay, resource depletion and injection attack [20]. | Supervised IDS for attack classification [21]. |
| | | Client-server model, Authenticating server's key with sensor data value [22]. |

### 4.3. Network Layer

This layer uses various technologies such as Radio Frequency Identification (RFID), Instrument flight rules (IFR), 3G, GSM, BLE, Universal Mobile Telecommunications System (UMTS), WiFi, ZigBee, etc for communicating with the devices. Communication in IoT devices occurs by routing and it is prone to various attacks [25]. Routing attacks involves spoofing, selective forwarding, altering routing paths or replaying packets, sinkhole, warm-hole etc. These attacks may lead to DoS threats. Table 3 shows some of the attacks against network or data link layer and its protective measures. While DoS or DDoS can be launched in Transport Layer also.

1) Eavesdropping: During transmitting the data from the sensor node to the gateway or server, the data is susceptible to hijack. An adversary can listen to the data and alter it from wireless channel [26]. An attacker detects information of the user and perceives the message-ID, timestamps; source and destination address which leads to a serious threat to privacy. Many solutions exist though, the latest framework for Eavesdrop resistance using Visible Light Communication (VLC) is a promising solution for IoT devices security [27].

2) Resource consuming attacks: Various attacks such as unfairness, collision and exhaustion attacks are included in this category. In an unfairness attack, the attacker tries to use whole services and resources of the application without considering the prerequisite it has [9]. Sometimes, this affects the network performance at the MAC layer. In a Collision attack, an attacker sends packets at the same frequency concurrently, which leads to collision and degradation of network performance. It manipulates frame header such that the checksum mismatch occurs, which leads to discarding of the data frames at the destination end. Exhaustion attack occurs when a channel is continuously active for long time to drain the battery power [9]. This kind of attacks lead to the failure in providing service and functionalities to end-users. These attacks can be mitigated using similar solutions of Battery Exhaustion and Flooding attacks in Transport Layer.

3) Grey-hole attack: In a multi-hop environment, the data transmission occurs from one node to another node in multiple steps [28]. In this process, the node forwards packets in the next hop to the destination (gateway). Before forwarding the packets, the attacker may misguide the route or inject malicious code to broadcast it further and initiate a routing loop. Such an activity is a Grey-hole attack in which the packets may loop infinitely deteriorating the performance of the network. The security mechanisms for such attacks [29] [30] [31] are explained in below subsections.

4) Sinkhole attack: In this type, a malicious node enchants with the neighbour nodes to create routes via malicious code. Once the attacker compromises the system, this attack creates an open door for other attacks [28]. It is very difficult to detect the sinkhole, selective forwarding and eavesdropping attacks in a network. Similar to this, in Sybil attack, a falsify node is present in the network with multiple fake identities deceiving the neighbouring nodes. Pretence, Masquerade and Replay attack mean the same. This attack also takes place in healthcare IoT devices, an illegitimate node behaves as a genuine node in the network, and it sends fake information to the remote area requesting treatment and an emergency team will respond to the non-existent patient [32]. This keeps the emergency staff busy, delaying and unattended to the real patients. A Denial-of-Service attack can be easily achievable by masquerade node. The captured data of masquerade node cause replay threat to the real-time IoT device application. Raza et al. [31] proposed an intrusion detection system for 6LoWPAN protocol targeting network routing attacks, sinkhole and selective-forwarding. The proposed IDS was developed using Contiki OS for IoT devices. It was successful to expose attacks in some situations but was unsuitable to smart home IoT devices.

5) Wormhole attack: wormhole attack is of similar kind in which an adversary receives packets from one location and then forwards and releases it to other location through a tunnel (wormhole). It is nearly impossible to detect or stop these types of attacks in a network using built-in security measures. Pongle et.al proposed an Intrusion Detection System to detect wormhole attacks in an IoT environment [30]. Nevertheless, the method is incapable of uncovering undefined cyber-attacks.

6) Denial-of-Service attacks: Data and network availability is a major security goal for IoT device applications. Mostly, in healthcare systems, threats of Denial-of-service are devastating because the devices and network need to be active and running all the time to monitor patients and to perform critical tasks [33]. Denial-of-Service and Distributed Denial-of-Service can affect the data, network performance and reliability of the whole network. There are two types of DoS attacks:

a) <u>Reprogramming Attack:</u> It refers to changing or modifying the source code. The application becomes inaccessible and sometimes it enters an infinite loop making the service/ resource unavailable to the requester. Robust authentication, strong access control mechanism and continuous monitoring is a recommendable solution for such attacks [34].

b) <u>Path-based DoS:</u> Numerous replay packets or spuriously injected packets overwhelms the sensor node by long-distance end-to-end communication path [35].

Researchers suggest a defensive approach based on the maximum magnitude of each middleware layer to handle such type of DoS/ DDoS attacks. The system checks for the number of requests sent to be under the predicted threshold capacity and if it exceeds, it triggers an alert to the network administrator [36] and blocks the request. Moreover, recent IDS approaches using Machine Learning (ML) and SDN proved efficient in blocking many DoS attacks [37] [38].

Table 3. Network layer attacks.

| Attack | Technique & Implications | Defence Mechanism |
| --- | --- | --- |
| Eavesdrop | The attack hijacks data during transmission [26] such that an adversary can listen or alter the data. | Innovative visible light communication (VLC) method based on channel correlation and error estimation [27]. |
| Resource Consuming Attacks | Unfairness, Collision and Exhaustion attacks [9]. Failure in providing services. | Symmetric encryption and layered security mechanism using TLS [39]. |
| Modification-type attacks (Routing attacks) | Grey hole, Sinkhole, Black hole and Wormhole attacks. [28].<br><br>Built-in security measures like authentication and access control cannot mitigate or detect such attacks. | IDS for sinkhole and selective-forwarding attacks [31]. |
| | | IDS to detect wormhole attacks in an IoT environment [30]. |
| | | Specification-based approach for the RPL protocol monitors network intrusions and malicious behaviour [29]. |
| Sybil attack | A node with false identity, DoS or replay threat [32]. | Host-based IDS using SDN blocks the victim device. SAAS model [40]. |
| Denial-of-Service Attack | DoS, DDoS, Denial of Sleep, SYN Flood, DNS Flood, Ping Flood, UDP Flood, and ICMP Broadcast [41]. | SDN architecture to identify DDoS, worm propagation and port scan [37]. IDS coupled provide better security. |
| | | Evasion attacks against ML IDS can be mitigated using Gradient-based approach [38]. |

It is difficult to mitigate the attacks discussed by traditional security measures and needs upgradation. We infer that the usual countermeasures involving basic mechanisms are ineffective. In most of the cases strong authentication, access control and monitoring systems are effective in identifying, mitigating and halting cyber-attacks. In addition, IDS is capable to detect most of the types of attacks in Perception, Transport and Network layer. The below section is elaborates and summarizes the potential security practices extracted from the above discussion.

## 4. SECURITY MEASURES

Numerous connected IoTs gives various decentralized ways for attackers or malware to enter. The high-security measures create a bottleneck for adaptability and make the device complex and in turn invite new security concerns [42]. IoT demands different customization for different purposes. The security incorporation should ensure the adaptability of the device and must be scalable with the addition of more devices to the network. The enhancement of the following security practices is required in IoT devices to ensure better protection and to ensure the security properties namely authentication, access control, confidentiality, integrity, non-repudiation.

### 4.1. Robust authentication mechanism

IoT devices has the feature of password authentication for accessing its services. Weak or default passwords, botnets, Trojans stealing passwords, dictionary and brute-force attacks are a point of high concern against authentication [43]. Nowadays, security specialists recommend integration of two methods for stronger authentication.

1) Biometric Authentication: Replacement of authentication process from password-based authentication to biometric authentication guarantees higher security, as it is robust against usual password cracking attacks. It involves bio-features of the authorized users such as face recognition, fingerprinting, eye recognition etc. Ruhul Amin et al. [44] proposed a biometric authentication protocol for IoT devices operating in a distributed cloud-computing domain to overcome vulnerabilities of cloud multi-server. Of course, biometric authentication have some issues such as cost and complexity of the algorithms used but many solutions exists in the literature for such loopholes.

2) Multi-factor Authentication (MFA): It involves multi-step authentication process: 2-step or 3-step, which includes a combination of knowledge-based (passwords), ownership-based (card), bio-based (fingerprint) features. One-time authentication requires two or three features (credentials) of the user, such as PIN and OTP for confirming a bank transaction. Biometric authentication integrated with MFA guarantees robust authenticity [45] in the current security implementations. As authentication is the primary requirement in smart devices, robust authentication mechanisms are recommendable for better protection.

### 4.2. A robust access control mechanism

Access control and data protection on low power IoT devices have become the need for protection against expanding cyberattacks. According to the research, Biometric access control is most favourable in IoTs. It takes the biological attribute of the individual for verification and identification. In this process, it compares the activities of the individual with the stored patterns in the system. This mechanism is vital to avoid host/ Internal-based attacks. To prevent unethical approaches for medical devices, a biometric-based two-level secure access control model is developed [46]. In this, the model converts the iris image to iris code. The verification of iris code is done by using hamming distance. It stores the master key in the system, employs less computation, and has a very small overhead. However, it involves a higher cost for biometric processing. To overcome this problem, many researchers have proposed advanced methods that minimize the cost of deploying. One such is framework has been developed using physical unclonable functions (PUFs) and hardware obfuscation by Nima et. al [47]. This method protects against access control circumvention and does not require key storage. This suggests that biometric or any other robust access control mechanism with less complexity guarantee security of IoT devices.

### 4.3. Software-defined networking (SDN)

Software-defined networking is the trending network security management in various application areas like business, smart homes and e-health systems. Any computer network consists of switches and routers as the main components. The important functions of switches/ routers are control plane and data plane. Control plane is responsible for where to send the traffic, whereas data plane forwards the traffic to a specific destination. In conventional networking, data plane and control plane are coupled. In SDN architecture, the control plane is separated from the data plane. A software-based entity, called controller, remotely controls the tasks of control plane [49]. The data plane executes in the hardware and control plane in the software and resides in a logically centralized way. SDN is capable to monitor network traffic and detect malicious activities. It identifies and isolates the compromised nodes from the rest of the network. Giotis et. al [37] used flow statistics in SDN architectures to spot abnormalities by using various ways such as launching a DDoS, worm propagation and port scan. It was efficient to detect attacks and does not cause overhead to the controller, but was not able to diagnose other attacks. However, SDN accompanied by an Intrusion Detection System is potential to identify or diagnose newer attacks [50] as per the latest research.

### 4.4. Intrusion detection system

Intrusion detection systems (IDS) is a program or algorithm, which tries to recognize malicious activities in a network. It also attempts to detect when a computer is under attack or an intruder is trying to compromise it. Besides, it identifies if a legitimate user is trying to escalate privileges or attempting to access unauthorized data or services. IDS has become an essential element for protecting the ICT infrastructure [50]. Nowadays, every network has IDS or IPS to detect and mitigate cyberattacks. According to the deployment model and data analysis, IDS is categorized as Network-based, Host-based or Application-based. In some contexts, system-based and application-based are considered as the two cases of Host-based IDS. Moreover, based on the technique / method used, IDS is categorized as Signature-based, Anomaly-based and Specification-based [51] [52]. An IDS system must distinguish attacks accurately, quickly and efficiently with less false alarms. Any IDS which identifies attacks accurately but takes a long time for detection is not suitable for current IoT networks [8]. Hence, it has become imperative to investigate a method that is capable to detect emerging attacks with less false alarms, which can handle a huge amount of data and take decisions quickly for real-time attack detection. The Signature-based system detects attacks based on signatures and known attack patterns but it is difficult to unmask known attack deviations or unknown attacks [53]. In literature, most of the implementations of IDS are rule-based which are inefficient in detecting novel attacks [32]. However, if the attack signatures database is up to date by adding new attack signatures every time, then this method is effective. Similarly, specification-based involves defining of rules by the administrator. In both these cases, the problem is the burden on the administrator to adapt to the changing number of devices and attacks. Anomaly Detection System detects deviations from a predefined normal behaviour but creates many false alarms for legitimate behaviours also when the user profile is complex and unknown. It is challenging to keep the IDS database up-to-date because of the heterogeneous network and changing environments such as network topology, servers, and several connected devices, communication protocols and open ports. To overcome this problem, the researchers are focusing on adaptable methods like Artificial Intelligence, Machine Learning and Deep Learning techniques [54].

### 4.4.1 Machine Learning IDS

This subsection provides details about the recent machine learning based Intrusion Detection Systems in IoTs. Mehdi et al. [40] proposed a host-based intrusion detection and mitigation system using OpenFlow protocol for security of smart home network. The scheme monitors the devices in home network to investigate the malicious activities and blocks the intruder to use the victim device once an intrusion is detection. The users in a smart home lack expertise in using security mechanisms, so Software Defined Networking (SDN) is employed in this model, providing Security as a Service (SaaS), such that a third party security specialist can monitor and take necessary actions when required. To avoid overburdening of nodes and communication network, host-based approach using filters is recommendable to monitor only suspicious nodes or malicious activities. In addition, the framework has the scalability to support heterogeneous new devices and technologies. The module consists of a database, which includes all the devices present in the smart home, their associated risks, and types of attacks and associated mitigation procedures for those. This model is based on Machine learning techniques which uses learned signature patters of known attacks. For this process, a sensor element gathers data traffic from suspicious nodes and send it to SDN controller. The captured traffic is transformed to service provider for feature extraction and to create predictive models of attacks. IoT Intrusion Detection and Mitigation (IDM) model uses linear regression and Software Vector Machine (SVM) to create a classification model, based on which attack is identified. Once an attack is detected, an alarm is raised, the victim node and attacker are identified and/or mitigation is done if measures are available in the database. This model was tested on a real IoT device, a smart lighting system and was proved efficient to detect the attack. The major disadvantage of this approach is that, each time a new device is added to the network, it has to be manually updated in the database. Only specific devices can be monitored and this approach is not feasible to investigate all devices in a home network. Moreover, in current zero-day attacks scenario, this approach is unsuitable as unknown attacks are not detected which may have devastating implications on the end-users.

Similarly, Kleber et al. [55] proposed a mechanism to overcome the huge number of IDS alerts, which are triggered in a conventional IDS system. This model is based on the fusion of various events, security logs and alerts and is not concerned with network traffic. The proposed scheme gathers raw data and change it in a standard normalized format. Then these normalized events are clustered into meta-sevents, to represent possible attack scheme more clearly when compared to the disconnected alerts. With this situational awareness, in the final state, the meta-events are classified using machine learning to categorize it as an attack or false alarm. SVM, Decision Tree and Bayesian Network have validated the classification scheme. This was tested using DARPA Intrusion Evaluation challenge [56] and SotM from the honeynet and the accuracy was in between 40 - 60% in attack detection with lesser false positive rates and was able to detect some of the newer attacks as well. However, this model was not been tested for current zero-day attacks and may not be feasible to detect multi-stage attacks. Various improvements are necessary in terms of security and complexity of the classification taxonomy of the approach.

Heena et al. [57] developed another machine learning approach for wireless sensor network security based on human immune system. This method intelligently detect anomalies by classifying the nodes into two categories: fraudulent or benevolent nodes. After which, the mechanism create virtual antibodies and depending on that, the gateway takes a decision whether or not to attack the fraudulent nodes. The model works similar to human immune system as a second line defense in the body. However, the actual implementation of the proposed mechanism was not provided. Likewise, Sara et al. [58] proposed an IDS Machine Learning based on feature selection and clustering algorithm incorporating filter and wrapper methods using linear correlation coefficient (FGLCC) algorithm, cuttlefish algorithm (CFA) and Decision Trees for classification. The authors verified the proposed method using KDD Cup 99 large data sets, which gave 95% detection rate.

### 4.4.2 Deep Learning IDS

Feature extraction and data classification has emerged as efficient techniques for IDS. However, most of the proposed approaches are inefficient when dataset is of large size. Kabir et al. proposed an Intrusion Detection System using Least Square Support Vector Machine (LS-SVM) in which the attack detection is done in 2 steps. In the first step, the entire dataset splits into subgroups such that they represent the whole dataset. In the second step, LS-SVM is applied to the proposed algorithm to determine intrusions. Various experiments using KDD 99 database proved it an efficient algorithm for intrusion detection [59]. The advantage of this method is that it supports static and incremental data also.

Papamartzivanos et al. [50] proposed an intelligent adaptive misuse Intrusion Detection System using Deep Learning. This method can adapt and sustain to various network environments with higher rates of attack detection. They used autonomic computing Self-Taught Learning method supported by MAPE-K model to assist IDS in new environments. This model is integrated with

MAPE-K method to create a framework for the autonomous and adaptive system. The model has been evaluated with various environmental changes and was capable to adapt with detection rate of approximately 73.3 % by not only detecting the attack but also categorizing it so that solution can be found easily. The benefits of deep learning methodologies is training the IDS based on the network activity in new environment.

From the above discussion, we deduce few quantitative metrics in Table 4, for an IDS system to determine its effectiveness.

Table 4. IDS Quantitative Metrics

| Metrics | Description |
| --- | --- |
| Coverage | The number and types of attacks that IDS can detect in a realistic environment. |
| Handling Traffic Bandwidth | The Ability of the IDS to handle High bandwidth traffic, block or resist traffic greater than the bandwidth of the channel. |
| Resisting attacks against IDS | Few attackers target IDS so that when the IDS is compromised, it becomes easier to attack the network and devices. Therefore, IDS must be capable to withstand attacks targeted against it. |
| Probability of Detection | It defines how accurately the system detects an intrusion. The approaches discussed have shown the accuracy up to 70 - 90%. Nevertheless, real-time networking demands higher accuracy in the detection of various attacks. |
| Probability of False Alarms [8] | Sometimes a non-attack activity is categorized as attack, and vice versa. These types of decisions of IDS may cause fatal implications. Latest Machine Learning and Deep Learning algorithms use Confusion matrix to correctly classify an event. |
| Ability to Detect Unknown Attacks [8] | The concern for Zero-day attacks are growing day by day. A system that is not able to detect unfamiliar attacks is inefficient. |
| Ability to Identify an Attack | How correctly a system is identifies an attack is important to take further actions by the network administrators. If the attack is categorized wrongly, for example, Wormhole attack is categorized as Grey hole; it may mislead the administrator by the risk level of the attack. |
| Ability to Determine Attack Success | The system must also be able to determine the status of the attack, such as, its success or failure, to what extent it is successful, and to what extent it damaged the resources. |
| Others | Other measurements include ease of use, deployment and maintenance. The IDS must meet resource requirements, performance and quality of service. |

### 4.5. Light-weight encryption

To secure the data at rest & data in transfer and to maintain confidentiality & integrity, encryption is necessary, which encapsulates the data in an unreadable format and it is decrypted at the receiver. Most common types of encryption algorithms are symmetric and as symmetric. Each of these approaches is capable to protect data against some attacks that another approach find it difficult. However, as-symmetric algorithms involve more processing which is unsuitable for IoT devices [60]. Currently, Physically Unclonable Function (PUF) is utilized for adding extra security hardware layer to protect against perception or physical layer attacks [61] and for end-points security. This method consumes lesser resources compared to other encryption algorithms. Whereas for communication security symmetric-key encryption is suitable, one such lightweight symmetric key encryption scheme has been developed [62], which provided very effective in securely transferring data in IoT networks.

The collaboration of the above-mentioned security properties is capable to defend against a maximum number of cyberattacks. In the below section, we provide an appropriate way of incorporating these security mechanisms in IoT architecture to acquire utmost protection.

## 5. SECURE IoT FRAMEWORK

From the above discussion, we have deduced that few of the security mechanism are significant to provide robust security for IoT devices. The summary of the findings from the study is shown in Figure 3, which is a proposal of secure architecture for IoT devices to protect it from malicious threats.

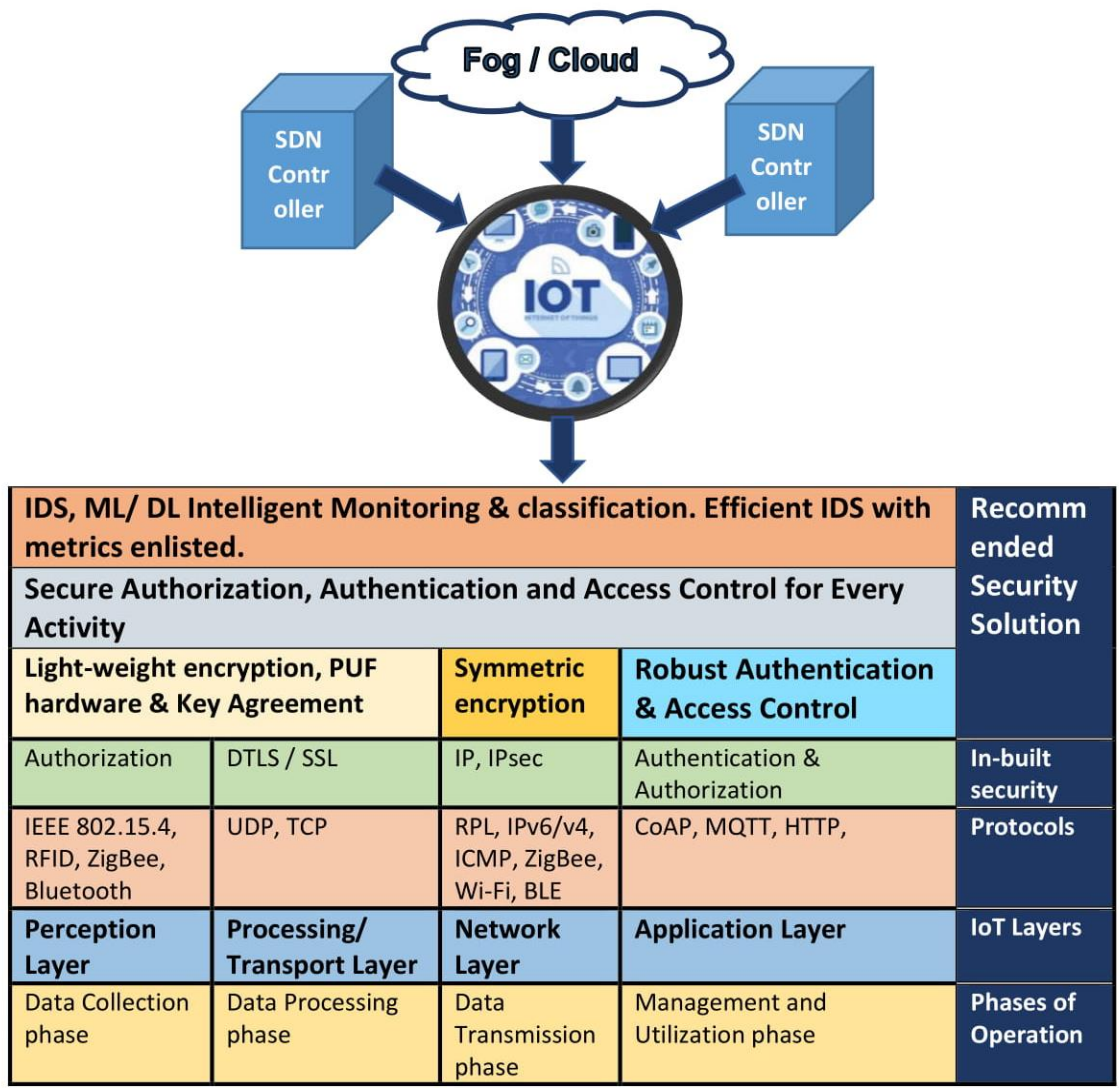

Figure 3. Secure IoT framework

While manufacturing the IoT devices, a standardized security layer needs to be included to provide basic security. The figure is explained from bottom Phases of operation to the topmost recommended security solution. The phases of operation and Layers of IoTs are already discussed in section 2. CoAP, MQTT and HTTP are frequently used protocols in Application Layer. Transport Layer consists of TCP and UDP protocols; similarly, the protocols of the other two layers are mentioned. Each layer has low-level in-built security, such as, to access application layer utilities the user has to authenticate his identity. At the network layer Internet Protocol Security (IPsec) is incorporated for secure communication. Similarly, at the transport layer, DTLS & Secure Software Layer (SSL) provides security and the perception layer customizes its activities based on the authorization process. The fundamental security controls in IoT architecture lag behind to defend it against current attacks.

According to this research, we have provided the necessary security mechanism at each layer to ensure optimal protection against cyberattacks. The security mechanisms discussed are placed appropriately to suit the layer requirements in the IoT architecture.

1) Robust authentication and access control: The application layer is protected against various attacks using robust authentication and access control mechanisms provided.

2) Symmetric Encryption & Light-weight cryptography: Transport and Network layer needs lightweight encryption combining the features of both symmetric and as-symmetric encryption to secure the ends and the data in transmission.

3) Secure authorization, authentication and access control: Every action at different layer needs to be checked for its authorization with proper access control.

4) Intelligent IDS: For the transport and network layer, regular monitoring can reduce the number of malicious intrusions. The attacks at the perception layer also can be mitigated by IDS monitoring and key agreement protocols.

5) Software Defined Networking: These days SDN architecture provides better security compared to other networking practices. Due to programming capability of SDN, secure and easily controllable network can be designed.

**5.1. Evaluation of Proposed Framework**

Putchala et al. [63] proposed a distributed multi-layered IDS architecture for IoT devices to ensure identifying malicious intrusions at each layer efficiently and accurately. The author suggested placement of Deep Learning based IDS at each layer for maximum coverage and better complexity. The implementation is tested and proved efficient. This suggest that IDS placed at all layers of IoT architecture guarantees better attack detection compared to one specific layer based IDS. Therefore, recommended security solution, IDS at all layers proved efficient. Lightweight encryption and authentication protocol proposed by researchers show that the protocol is protected against possible security threats [44]. In similar way, a robust cryptographic technique can be incorporated and tested along with the IDS system. The other mechanisms can be tested solely or in combination of all mechanisms to validate its effectiveness. In future, we aim to test all the mechanisms suggested and prove its effectiveness in terms of complexity, resource constraint requirements and various features.

The following are the metrics based on which the proposed framework has to be evaluated.

1) Processing response time: Performance of IoT device in a real environment after implementing all the recommended security mechanisms.

2) Resource consumption: The level of resources consumed, such as battery power, processing power and memory used.

3) Attacks mitigation: The number of attacks that are mitigated after implementing the framework, the accuracy of attack detection.

4) Scalability: The amount of data that is easily being processing without overwhelming device.

Other metrics are also included in future during the implementation of the framework and compare its effectiveness with the recently proposed secure IoT architectures. The framework needs to be tested by launching real-time attacks against the IoT device to ensure its deployment in current IoT devices.

## 6. CONCLUSION

With the advancement of heterogeneous smart IoT devices, the concern for security is increased. In this paper, we have provided various types of attacks against IoTs and their protective measures. Certainly, many other attacks are launched against IoT layers, but the attacks discussed here are recurrent and devastating. We have learnt that securing the endpoints, network monitoring and the protecting data in the transfer is mandatory, to detect and prevent malicious activities in IoTs. Thus, we have proposed a framework incorporating Robust Authentication, Robust Access Control, Lightweight cryptography and Intrusion detection system, to secure data in transfer, sensitive stored data, settings & privileges. SDN is a trendy networking paradigm to securely control the whole network using programming. Machine Learning and Artificial Intelligence (AI) is an emerging field for IDS, which allows a system to learn, deduce and decide based on cognitive functions of pattern recognition and computational learning theory without any programming. In addition, the metrics used by IDS to identify different attacks against IoT layers are provided. The proposed framework is a valuable contribution to the IoT architecture due to its holistic approach of combination of various potential security mechanism. In future, this research aims to implement the framework and validate its effectiveness in terms of security, performance and usability.